\documentclass[prl,twocolumn,tightenlines,superscriptaddress,nofootinbib,
showpacs]{revtex4}
\usepackage{float}
\usepackage{amssymb,latexsym}
\usepackage{amsmath,amsbsy,bbm}
\usepackage{epsfig,bm}
\usepackage{graphicx,comment}
\usepackage{color}
\begin{document} 

\title{A New Method of Calculating the Spin-Wave Velocity $c$
of Spin-1/2 Antiferromagnets With $O(N)$ Symmetry in a Monte Carlo Simulation}

\author{F.-J. Jiang}
\email[]{fjjiang@ntnu.edu.tw}
\affiliation{Department of Physics, National Taiwan Normal University,
88, Sec.4, Ting-Chou Rd., Taipei 116,
Taiwan}

\vspace{-1cm}
  
\begin{abstract}
Motivated by the so-called cubical regime in magnon chiral perturbation theory, we 
propose a new method to calculate the low-energy constant, namely the spin-wave 
velocity $c$ of spin-1/2 antiferromagnets with $O(N)$ symmetry in a Monte Carlo simulation. 
Specifically we suggest that $c$ can be 
determined by $c = L/\beta$ when the squares of the spatial and temporal 
winding numbers are tuned to be the same in the Monte Carlo calculations. 
Here $\beta$ and $L$ are the inverse temperature and the box size used in the simulations
when this condition is met. We verify the validity of this idea
by simulating the quantum spin-1/2 XY model. The $c$ obtained by using the squares of winding numbers
is given by $c = 1.1348(5)Ja$ which is consistent with the known values of $c$ in the literature. 
Unlike other conventional approaches, our new idea provides a direct method
to measure $c$. Further, by simultaneously fitting our Monte Carlo data of susceptibilities $\chi_{11}$
and spin susceptibilities $\chi$ to their theoretical predictions from magnon chiral perturbation
theory, we find $c$ is given by $c = 1.1347(2)Ja$ which agrees with the one we obtain by the 
new method of using the squares of winding numbers. The low-energy constants magnetization density ${\cal M}$ and
spin stiffenss $\rho$ of quantum spin-1/2 XY model are determined as well 
and are given by 
${\cal M} = 0.43561(1)/a^2$ and $\rho = 0.26974(5)J$, respectively.
Thanks to the prediction power of magnon chiral perturbation theory which puts
a very restricted constraint among the low-energy constants for the model considered 
here, the accuracy of ${\cal M}$ we present in this study is much precise
than previous Monte Carlo result.

\end{abstract}

\maketitle

\section{Introduction}
During the last twenty years, models with $O(N)$ symmetry which are relevant to antiferromagnets
have drawn a lot of attention. 
In particular spin-1/2 Heisenberg-type models have been 
studied in great detail both analytically and numerically 
because it is believed that these models
are the correct models to describe the undoped precursors of
high $T_c$ cuprates (undoped antiferromagnets).
Beside their phenomenological importance, these $O(N)$ models 
for antiferromagnets 
are interesting from theoretical perspective as well.
Further, because of the availability of efficient Monte Carlo algorithms and 
increasing computing power, the physics of these models 
has been investigated with unprecedented numerical 
accuracy \cite{Bea96,Sandvik97,Bea98,Sandvik99,Kim00,Wang05,Jiang08,Alb08,Wenzel09,Jiang09.1}.
For instance, using a loop algorithm as well
as finite-volume and -temperature predictions from magnon chiral
perturbation theory, the corresponding low-energy constants, namely
the staggered magnetization
density ${\cal M}_s$, the spin stiffness $\rho_s$ and the spin-wave velocity
$c$ of spin-1/2 Heisenberg model on the square lattices are determined with high accuracy and
are in agreement with experimental results \cite{Wie94}.
Because the properties of these models with $O(N)$ symmetry 
are well-studied,  
they are particular suitable for exploring any new idea.

In analogy to chiral perturbation theory for the pions
in QCD, a systematic low-energy effective field theory
for the magnons in an antiferromagnet exists as well and
is called magnon chiral perturbation theory \cite{Cha89,Neu89,Has91}. 
Low-energy effective field theories are based on symmetry constraints 
of the underlying models and are universally applicable. 
Results obtained by effective field theories are exact, order
by order in a systematic low-energy expansion. Material
specific properties enter the effective Lagrangian in the
form of a priori undetermined low-energy parameters, like
the spin stiffness $\rho$ ($\rho_s$) or the spin-wave velocity $c$. Once
the numerical values of these low-energy constants are determined,
either by Monte Carlo simulations or experimental data, the
low-energy physics of the underlying models are completely
determined. Since the low-energy physics of the underlying model
only depends on the corresponding low-energy constants,
it is important to determine these low-energy constants
as precise as possible. From theoretical perspective, to
determine the numerical values of these low-energy constants 
are important as well. For instance, by simulating spin-1/2
Heisenberg model with an external staggered field on an
exactly cubical space-time box (which requires a very precise value of
$c$), in addition to being able to determine the numerical values of ${\cal M}_s$
and $\rho_s$, such investigation also provides a good opportunity to 
exam the validity of predictions from the corresponding 
low-energy effective field theories \cite{Goe91,Goe91a}.

For (sub-lattice) magnetization density and spin stiffness, one can directly measure the related
observables and then use experimental finite lattice extrapolation formulae
to obtain the bulk values of these 2 low-energy constants. On the
other hand, the spin-wave velocity $c$ is always determined in a less direct manner conventionally.
Motivated by the so-called cubical regime (defined later) in magnon
chiral perturbation theory \cite{Has93}, we propose a new method to calculate
the spin-wave velocity $c$ of spin-1/2 antiferromagnets with $O(N)$ symmetry
in a Monte Carlo simulation\footnote{This method was implicitly used in 
\cite{Ger09} for the study of constraint effective potentials. Here we carry out quantitative investigation to verified the validity 
of this method.}. Specifically, we propose that
$c$ can be calculated by $c = L/\beta$ when the squares of
spatial and temporal winding numbers are tuned to be the same.
Here $L$ and $\beta $ are the spatial box size and the inverse temperature
used in the simulations when above condition is met. Since this method 
allows one to measure $c$ in a
direct manner, the result is more accurate than other methods. 
Indeed as we will demonstrate later, for quantum spin-1/2 XY model,
the numerical value of
$c = 1.1348(5)Ja$ we obtain using the new idea
is of high precision and is consistent with the known Monte Carlo results in the literature 
as well \cite{San00}.
Further, by simultaneously fitting our Monte Carlo data of susceptibilities $\chi_{11}$
and spin susceptibilities $\chi$ to their theoretical predictions from magnon chiral perturbation
theory \cite{Has93}, we find $c$ is given by $c = 1.1347(2)Ja$ 
which is consistent with the one we obtain by the new method.
These results confirm the validity and usefullness of our new method.
Additionally other two low-energy constants, namely the magnetization density ${\cal M}$
and spin stiffness $\rho$ of spin-1/2 XY model are calculated with high accuracy and are much precise
than previous Monte Carlo estimates \cite{San00}.


The remaining of this paper is organized as follows. After a brief 
introduction to our motivation of this study, we summarize
the model and observables investigated here. Follows that
we review the corresponding effective field theory predictions 
relevant to our study. Then we present
our numerical results. In particular we demonstrate the
validity of the method we used in our simulations to determine
the low-energy constant $c$. Along the verification,
the low-energy constants ${\cal M}$ and $\rho$ 
are also calculated. Finally a section is 
devoted to the conclusion of our investigation.
 

\section{Microscopic Models and Corresponding Observables}
The quantum XY model we consider in this study is defined by the Hamilton 
operator
\begin{eqnarray}
\label{hamilton}
H = \sum_{\langle i,j \rangle} J \Big[\, S^1_iS^1_j+ 
S^2_i S^2_j\,\Big],
\end{eqnarray}
where $S_i^1$ and $S_i^2$
are the first and second components of a spin-
1/2 operator at site $i$, and $i$, $j$ denotes a pair of nearest
neighbor sites on a square lattice. Further, 
$J$ in eq.~(\ref{hamilton}) is the antiferromagnetic coupling.
A physical quantity of central interest is the susceptibility 
which is given by
\begin{eqnarray}
\label{defstagg}
\chi_{11} 
&=& \frac{1}{L^2} \int_0^\beta dt \ \frac{1}{Z} 
\mbox{Tr}[M^1(0) M^1(t) \exp(- \beta H)].
\end{eqnarray}
Here $\beta$ is the inverse temperature, $L$ is the spatial box
size, $Z = \mbox{Tr}\exp(- \beta H)$
is the partition function and $M^1 = \sum_x S^1_x$  
is the first component of magnetization.
Another relevant quantity is the spin susceptibility which is 
given by
\begin{eqnarray}
\label{defuniform}
\chi 
&=& \frac{1}{L^2} \int_0^\beta dt \ \frac{1}{Z} \mbox{Tr}[M^3(0) M^3(t)
\exp(- \beta H)],
\end{eqnarray} 
here $M^{3} = \sum_{x}S^3_{x}$. Both $\chi_{11}$ and $\chi$ can be measured very 
efficiently with the loop-cluster algorithm using improved estimators 
\cite{Wie94}. In particular, in the multi-cluster version of the algorithm the 
susceptibility is given in terms of the cluster sizes $|{\cal C}|$, 
i.e.
$\chi_{11} = \frac{1}{ \beta L^2 } \left\langle \sum_{\cal C} |{\cal C}|^2 
\right\rangle$.
Similarly, the spin susceptibility
$\chi = \frac{\beta}{ L^2 } \left\langle W_t^2 \right\rangle =
\frac{\beta}{ L^2 } \left\langle \sum_{\cal C} W_t({\cal C})^2 
\right\rangle$
is given in terms of the temporal winding number 
$W_t = \sum_{\cal C} W_t({\cal C})$ which is the sum of winding numbers
$W_t({\cal C})$ of the loop-clusters ${\cal C}$ around the Euclidean time 
direction. Finally, the spatial winding numbers are defined by 
$W_i = \sum_{\cal C} W_i({\cal C})$ with $i \in \{1,2\}$.

\section{Low-Energy Effective Theory for Magnons}
Due to the spontaneous breaking of the global $O(2)$ symmetry, 
the low-energy physics of antiferromagnets with 
an $O(2)$ symmetry is governed by
one massless Goldstone boson. 
Detailed calculations of a variety of 
physical quantities for the spin-1/2 antiferromagnets with $O(N)$ symmetry
including the NNLO contributions 
have been carried out in \cite{Has93}. Here we only quote the results that are 
relevant to our study. The aspect ratio of a 
spatially quadratic space-time box with box size $L$ is characterized 
by $l = (\beta c /L )^{1/3}\,,$ with which one distinguishes cubical space-time
volumes with $\beta c \approx L$ from cylindrical ones with $\beta c \gg L$.
The $c$ appearing above is the low-energy constant spin-wave velocity. 
In the cubical regime, the volume- and temperature-dependence of the 
susceptibility is given by
\begin{eqnarray}
\label{chiscube}
\chi_{11} &=& \frac{{\cal M}^2 L^2 \beta}{2} 
\left\{1 + \frac{c}{\rho_s L l} \beta_1(l) \right. \nonumber \\
&+&\left.\frac{1}{2}\left(\frac{c}{\rho_s L l}\right)^2 
\left[\beta_1(l)^2 \right. \right. \nonumber \\ 
&+&\left. \left.
\beta_2(l)\right] + O\left(\frac{1}{L^3}\right) \right\},
\end{eqnarray}
where ${\cal M}$ is the magnetization density and $\rho$ is
the spin stiffness. Further 
the spin susceptibility in the cubical regime takes the form
\begin{eqnarray}
\label{chiucube}
\chi &=& \frac{\rho}{ c^2} 
\left\{1 + O\left(\frac{1}{L^3}\right) \right\}.
\end{eqnarray}
In eq.~(\ref{chiscube}), the 
functions $\beta_i(l)$, which only 
depend on $l$, are shape coefficients of the space-time box defined in 
\cite{Has93}.
Finally, in the cylindrical regime, the temperature- and volume-dependence 
of the   
spin susceptibility $\chi$ at very
low temperature, namely when the condition $L^2\rho/(\beta c^2) \ll 1$ is satisfied, is given by
\begin{eqnarray}
\label{chiucylin}
\chi = 2\frac{\beta}{L^2}\exp \Bigg[-\frac{1}{2}\frac{c^2\beta}{\rho L^2}\Bigg]
.\end{eqnarray}


\section{The Determination of the Low-Energy Constants}
Conventionally the spin-wave velocity $c$ is determinated indirectly in a Monte
Carlo simulation. For example, magnon chiral perturbation theory predicts that
the finite-volume dependence of the ground state 
internal energy density is given by \cite{Has93}
\begin{equation}
\label{genergyd}
e_0(L) = e_0 + 0.7188725 c / L^3 + O(1/L^5)  
.\end{equation}
By fitting the related Monte Carlo data to above equation, $c$ can be obtained
from the coefficient associated with the term of $1/L^3$ in 
eq.~(\ref{genergyd}).
Further, $c$ can be 
calculated by the standard hydrodynamic relation
$\chi =\rho/c^2$ as well. Notice for both methods mentioned
above, extrapolations to infinite volume limit are necessary 
in order to obtain the numerical value of $c$.
In other word, $c$ is determined indirectly and the step of 
extraoplations will introduce systematic 
uncertainties
into its numerical value. Because of this, here we propose a new method 
to calculate the numerical value of $c$ for the spin-1/2
antiferromagnetic models with $O(N)$ symmetry directly in a Monte Carlo simulation. 
Our new idea is motivated
by the so-called cubical regime in magon chiral perturbation 
theory \cite{Has93}. Specifically
in magon chiral perturbation theory, an exactly cubical space-time box is obtained 
through the condition $\beta c = L$, here again $\beta$ and $L$ are
the inverse temperature and the spatial box size, respectively. Further, in a Monte Carlo calculation, if one simulates
the system in an exactly cubical space-time box, then the average of the square of spatial 
winding numbers $\langle W^2 \rangle = 1/2(\langle W_1^2 \rangle + \langle W_2^2 \rangle)$ 
and the square of temporal winding number $\langle W_t^2\rangle$
should be the same. Once the condition $\langle W^2 \rangle = \langle W_t^2 \rangle$ 
is met, $c$ can be determined by $c = L/\beta$. In practice, to employ this new method to calculate
$c$, for a given box size $L$, one varied $\beta$ until the condition 
$\langle W^2 \rangle = \langle W_t^2 \rangle$ is reached.
We would like to emphasize that the method we propose here to determine the low-energy
constant $c$ applies to
any quantum spin-1/2 antiferromagnetic system with a spontaneous symmetry breaking from 
a global $O(N)$ symmetry to its $O(N-1)$ subgroup.
Notice since the $c$ in the criterion of an exactly cubical space-time box, 
namely $\beta c = L$ is its bulk value,
one would expect the $c$ calculated by this new method suffers very mild 
finite lattice effects. Indeed,
as we will demonstrate shortly, for the quantum XY model considered in this study, th nunerical value of $c$
obtained by the new method is saturated to its bulk value even at 
$L = 24a$.

\begin{figure}
\begin{center}
\vbox{
\includegraphics[width=0.345\textwidth]{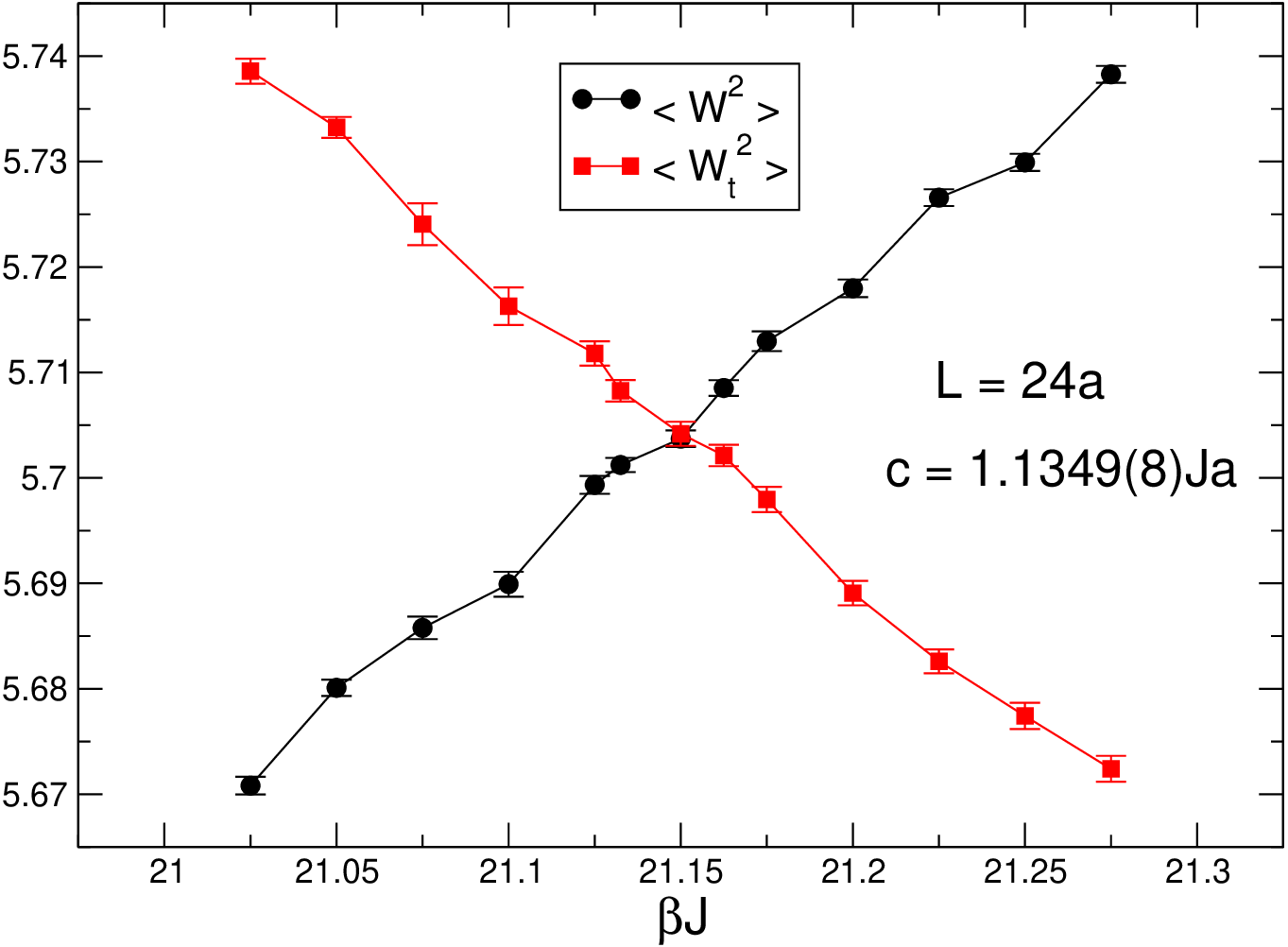}\vskip0.85cm
\includegraphics[width=0.345\textwidth]{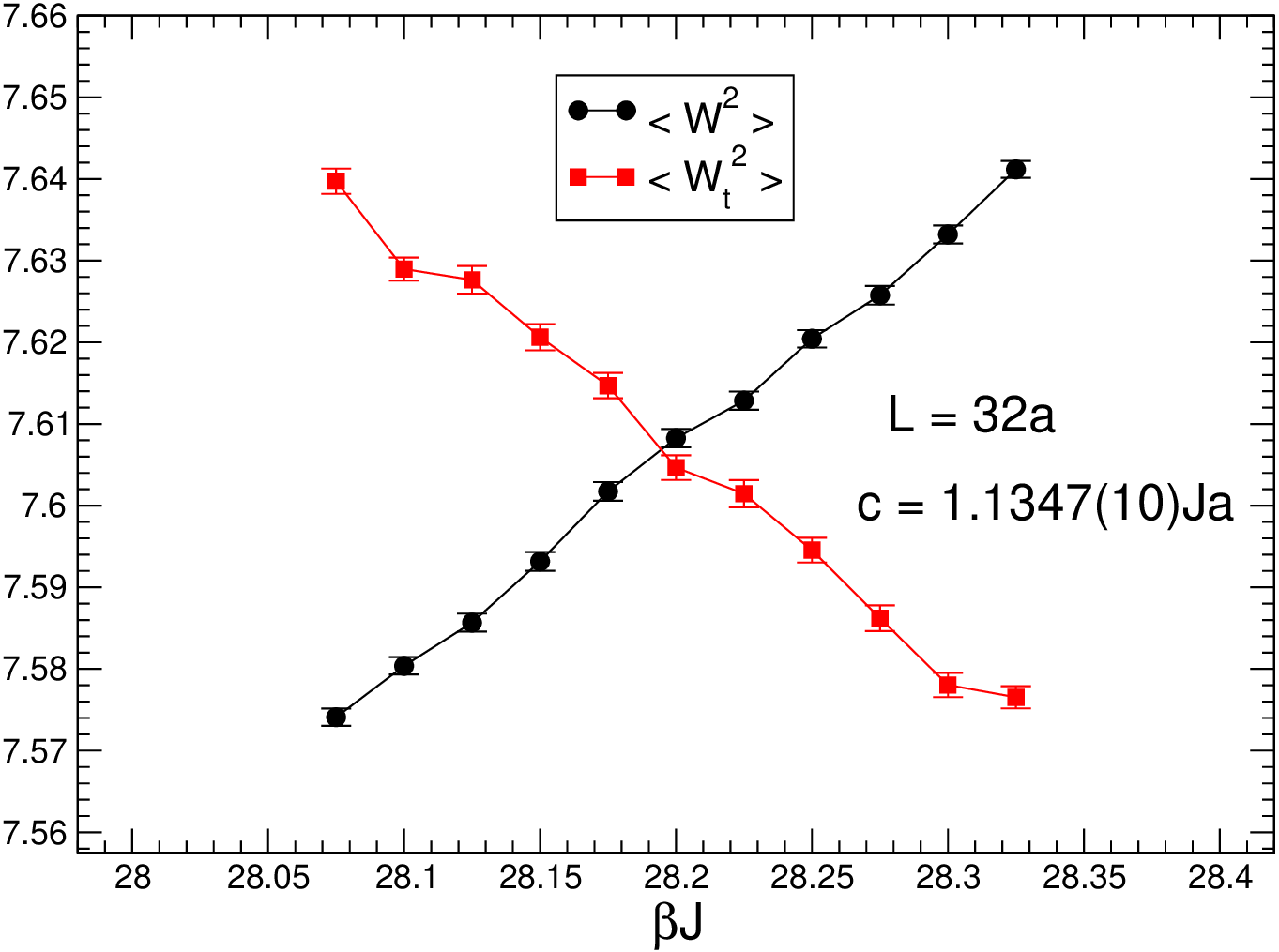}\vskip0.85cm
\includegraphics[width=0.345\textwidth]{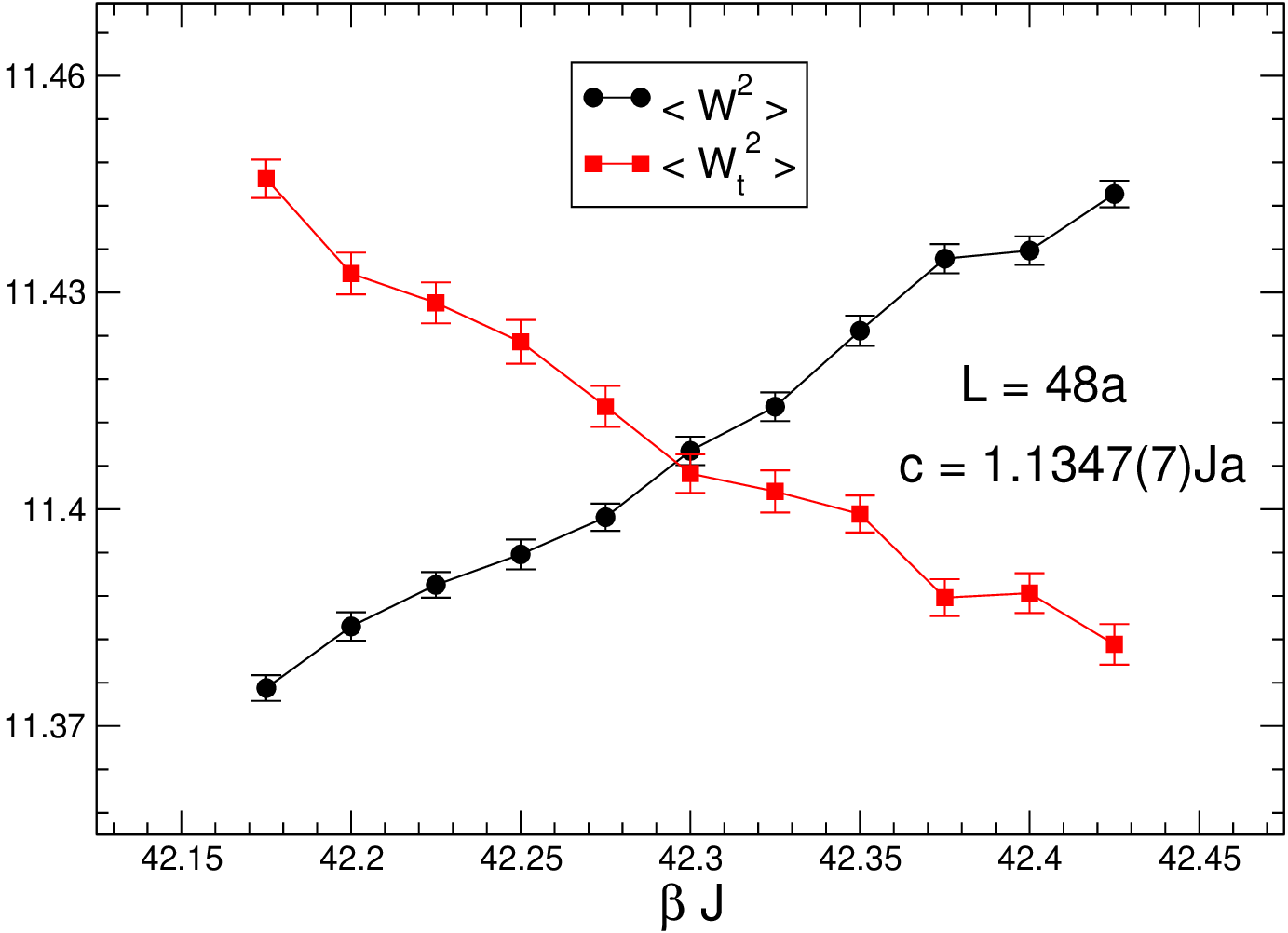}
}
\end{center}
\caption{The determination of $c$ using the squares of spatial and temporal winding numbers.}\vskip0.3cm
\label{fig1}
\end{figure}

To verify the validity of the new method we propose here to calculate $c$ in Monte Carlo simulations, 
we have carried out several simulations using a continuum-time loop algorithm
with $L = 24a,\,32a$ and $L = 48a$. Further, by tuning the inverse temperature
$\beta$ for each simulations to reach an exactly cubical space-time box, the numerical
values for $c$ determined from these simulations with $L = 24a,\,32a,\,48a,\,$ 
are given by $c = 1.1349(8)Ja$, $c = 1.1347(10)Ja$ and $c = 1.1347(7)Ja$, respectively. 
Figure \ref{fig1} demonstrates the
results of such calculations. The 3 values for $c$ obtained at
different box sizes are consistent with each other and agree with earlier Monte Carlo
result of $c$ as well \cite{San00}. This provides a convinving evidence to support the validity of our
new method of calculating $c$ from the squares of spatial and temporal winding numbers.      
By a weighted average over these values of $c$ determined at different box sizes, the
final result of the numerical value for the low-energy constant $c$ in this study 
obtained by the new method is given $c = 1.1348(5)Ja$. 
We have additionally carried out
simulations with $L=72a$ and $\beta = 63.450099141/J$ (which corresponds to $c=1.13475 Ja$).
The $\langle W^2 \rangle$ and $\langle W_t^2 \rangle$ obtained from these new runs 
are given by $17.1150(95)$ and $17.109(16)$, respectively.
This result implies that the method of calculating $c$ through the squares of spatial and temporal winding numbers
indeed suffers very mild finite volume effects, at least for the model considered in this study. 
We notice that the $\rho$ corresponding to these new runs
is given by $\rho = 0.26973(15) J$ which is statistically consistent with 
$\rho = 0.26975(8) J$ calculated at $L = 24a$.

\begin{figure}
\begin{center}
\includegraphics[width=0.345\textwidth]{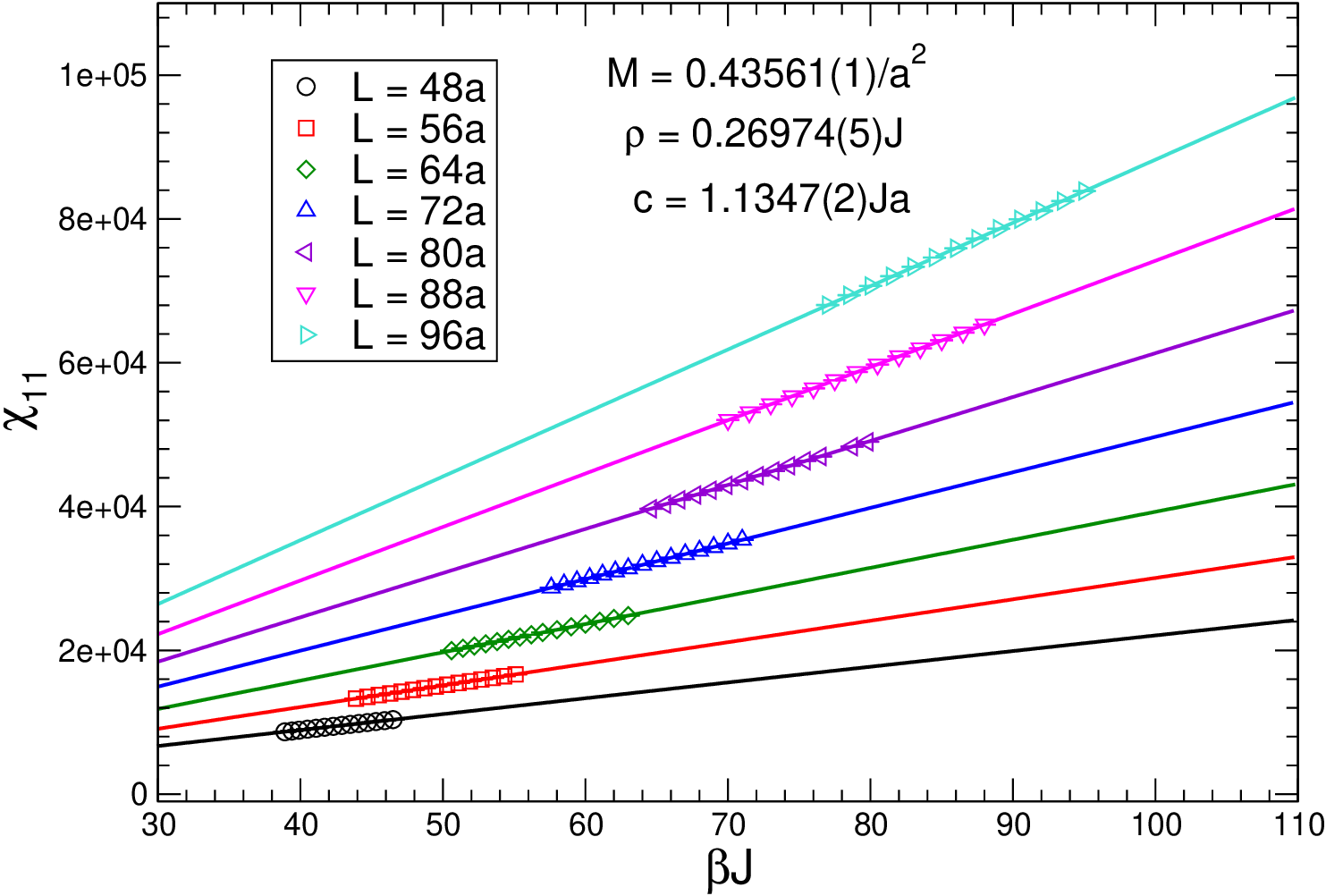}
\end{center}
\caption{Result of fitting the data points of $\chi_{11}$ obtained in the cubical regime
to their magnon chiral perturbation theory prediction. Some data points
are omitted for better visibility.}
\label{fig2}
\end{figure}

\begin{figure}
\begin{center}
\includegraphics[width=0.345\textwidth]{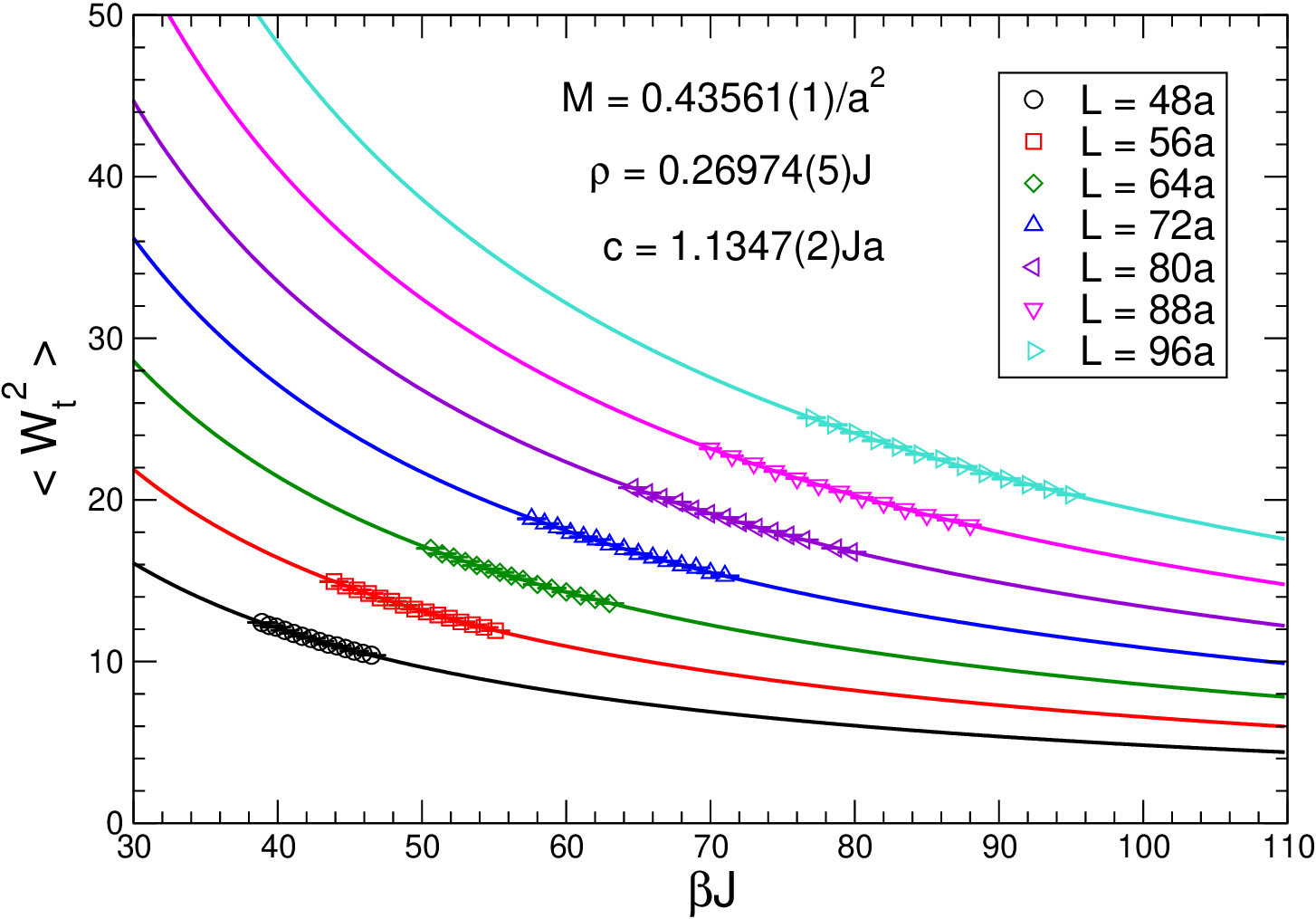}
\end{center}
\caption{Result of fitting the data points of $\langle W_t^2 \rangle$ obtained in the cubical regime
to their magnon chiral perturbation theory prediction. Some data points
are omitted for better visibility.}\vskip0.2cm
\label{fig3}
\end{figure}

\begin{figure}[H]
\begin{center}
\includegraphics[width=0.345\textwidth]{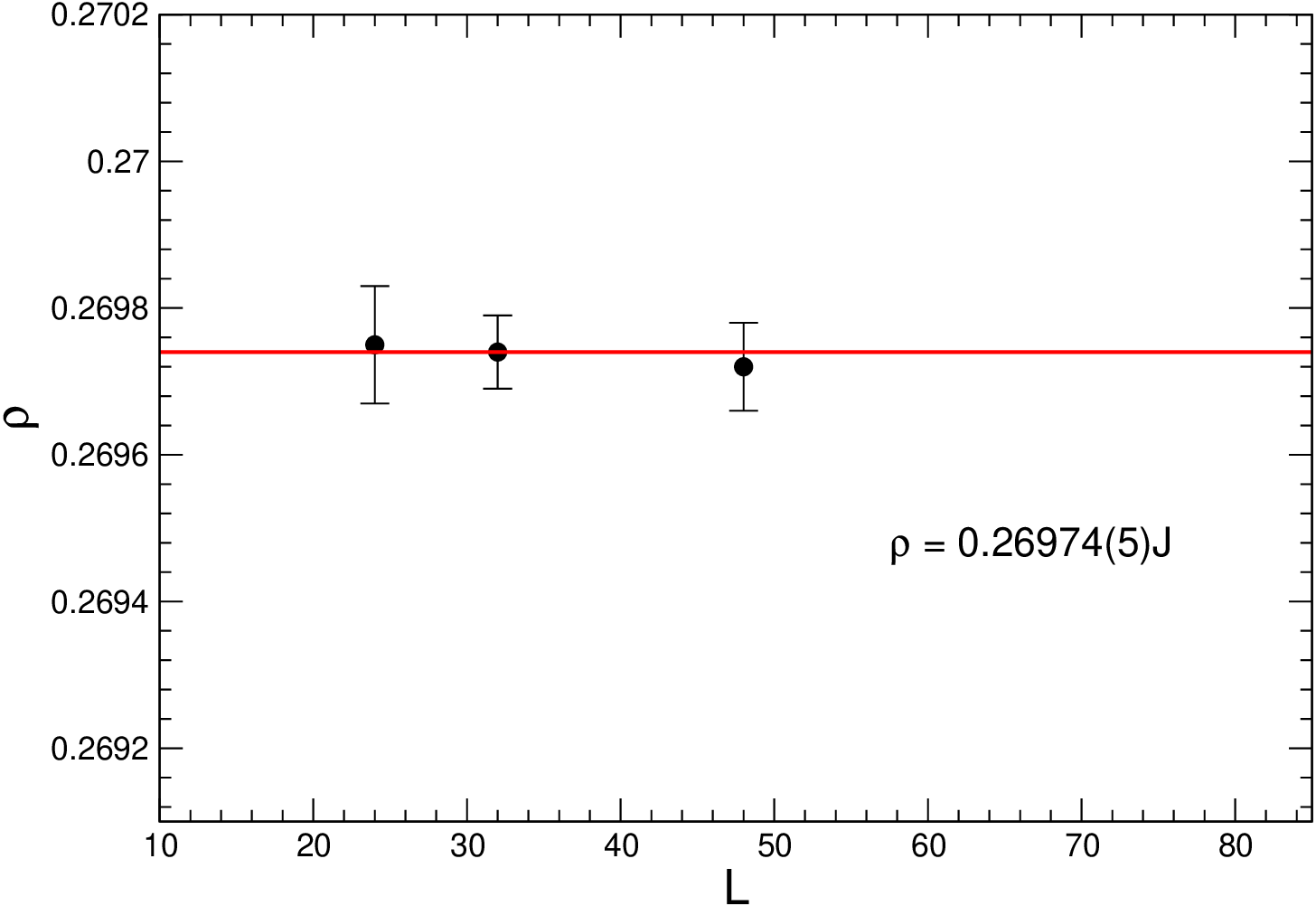}
\end{center}
\caption{Result of fitting the data points of $\rho$ obtained in the cubical regime
to a constant.}
\label{fig4}
\end{figure}

Another way to exam whether the new idea of calculating $c$ through the squares of
winding numbers is quantitatively correct is to extract $c$ by fitting Monte Carlo data of $\chi_{11}$ and
$\chi$ in the cubical regime to their predicted volume- and temperature-dependence, namely eqs.~(\ref{chiscube}) 
and (\ref{chiucube}), respectively. However notice since
$c$ always appears as the quantity $c/\rho$ or $\rho^2/c$ in eqs.~(\ref{chiscube}) and (\ref{chiucube}),
$c$ and $\rho$ are highly correlated and it would be a challenge to extract 
$c$ accurately from the fits. Fortunately we observe from our Monte Carlo
data that the observable $\langle W^2 \rangle$ which is exactly $\rho \beta$
when $L \rightarrow \infty$ already reaches a constant for $L \ge 24a$. 
Hence in our simultaneous fits,
we also include an improved estimate of the 
$\langle W^2 \rangle /\beta$ data obtained earlier
when determining $c$ using the new method and fit these data points 
to a constant. By simultaneously fitting cubical regime data of
$\chi_{11}$ and $\chi$ with $L \ge 24a$ as well as the data of 
$\langle W^2 \rangle /\beta$ to their predicted volume- and 
temperature-dependence formulae, we arrive at
${\cal M} = 0.43561(1)/a^2$, $\rho = 0.26974(5)J$ and $c = 1.1347(2)Ja$
with a $\chi^2/{\text{d.o.f.}} \sim 1$. 
The results of the fit
are shown in figures \ref{fig2}, \ref{fig3} and \ref{fig4}. 
The value of $c$ calculated from the chiral fits
agree nicely with the one detemined using the new method. 
Using large volume data points ($L \ge 40a$) for the fits leads to consistent
results. This in turn proves the
quantitative correctness of the new method of determining $c$ using the squares of winding numbers. Notice the values
for ${\cal M}$ and $\rho$ we obtain are consistent with the known 
values from Monte Carlo simulations \cite{San00}. They are in good 
agreement with the related results from 
series expansion and spin-wave calculations
in the literature as well \cite{Ham91}. Further, the numerical values of these low-energy constants
we obtain are much precise than those calculated in earlier Monte Carlo 
study. Finally
using the values for $\rho$ and $c$ determined in the cubical regime as well
as eq.~(\ref{chiucylin}), we have compared the theoretical prediction and Monte Carlo data
for $\chi$ in the cylindrical regime. The result of such comparsion is shown in figure
\ref{fig5}. Considering the fact that there is no free parameter, 
the agreement demonstrated in figure \ref{fig5} is reasonably good. 
All the results presented in this study also provides
a strong support for the prediction power and quantitative correctness of 
magnon chiral perturbation theory in understanding the low-energy physics
of the underlying model.

\section{Conclusions}
\begin{figure}[H]
\begin{center}
\includegraphics[width=0.345\textwidth]{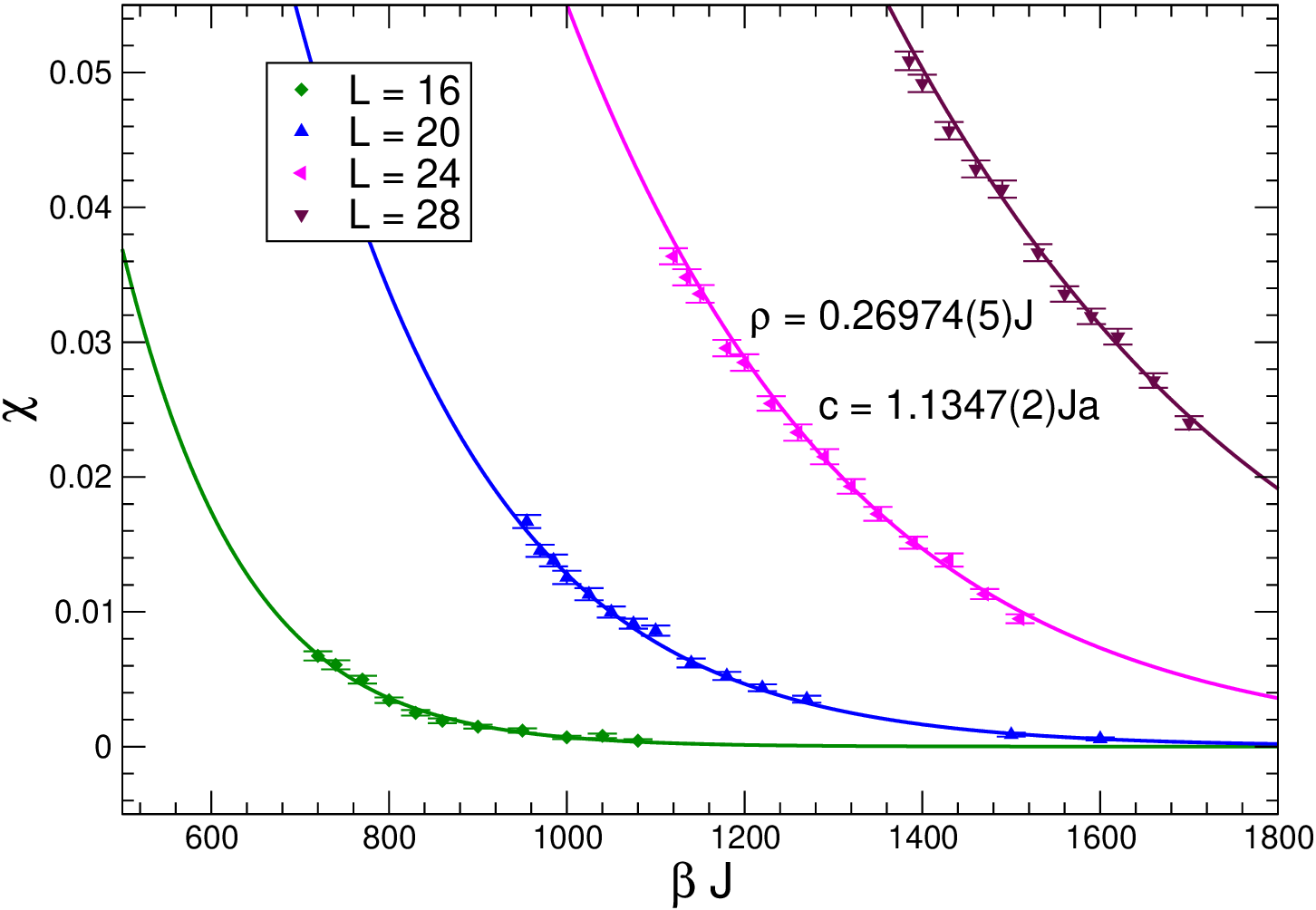}
\end{center}
\caption{Comparison between theoretical prediction and Monte Carlo data for 
$\chi$ in the cylindrical regime.}
\label{fig5}
\end{figure}

In this paper we have proposed a new method to calculate the low-energy constant, 
namely the spin-wave velocity $c$ for general 
antiferromagnetic spin systems with a spontaneous symmetry breaking 
from a global $O(N)$ symmetry to its $O(N-1)$ subgroup
using the squares of spatial and temporal winding numbers.  
We have demonstrated 
the validlity of this method by simulating the quantum spin-1/2 XY model. The numerical value of $c$ we calculate
with the new idea is given by $c = 1.1348(5)Ja$ which is consistent with the known Monte 
Carlo result in the literature. By fitting our Monte Carlo data of $\chi_{11}$ and $\chi$
to their volume- and temperature-dependence predictions from magnon chiral perturbation theory,
we reach ${\cal M} = 0.43561(1)/a^2$, $\rho = 0.26974(5)J$ and $c = 1.1347(2)Ja$. The value
of $c$ obtained from the fit is consistent with the one determined by the new method using winding numbers 
squared. This supports strongly the quantitative correctness of the new method we propose here
to calculate the low-energy constant spin-wave velocity $c$ in Monte Carlo simulations. The idea of using winding number squared is 
simple, but very powerful and requires
moderate computational effort to obtain a very precise numerical value for $c$. Finally thanks to the robustness nature and
prediction power of magnon chiral perturbation theory which puts very restricted
constraints on the low-energy constants and observables considered here, we are able to
fit simultaneously our finite temperature data points to their predicted formulae from magnon
chiral perturbation theory and obtain very accurate values for ${\cal M}$, $\rho$ and $c$. 
The agreement between theoretical prediction and
Monte Carlo results of $\chi$ at very low temperature shown in figure \ref{fig5}
is remarkble as well considering the fact that there is no free parameter in obtaining 
figure \ref{fig5}.

\section{Acknowledgements}
The simulations
in this study were done based on the loop algorithms available in 
ALPS library \cite{Troyer08}. We would like to thank U. Gerber and W.-J. Wiese
for useful correspondence. Partial support from NCTS (North) is acknowledged.

\vskip1.0cm
  
\end{document}